\documentclass[10pt,twocolumn,letterpaper]{article}

\usepackage{wacv}
\usepackage{times}
\usepackage{epsfig}
\usepackage{graphicx}
\usepackage{amsmath}
\usepackage{amssymb}
\usepackage{url}
\usepackage{graphicx}
\usepackage{array}

\wacvfinalcopy 

\def\wacvPaperID{214} 

\ifwacvfinal\fi
\begin{document}

\title{Crime Mapping from Satellite Imagery via Deep Learning}

\author{Alameen Najjar \hspace{1cm} Shun'ichi Kaneko \hspace{1cm} Yoshikazu Miyanaga \\
Graduate School of Information Science and Technology, Hokkaido University, Japan\\
\tt\small najjar@hce.ist.hokudai.ac.jp, \{kaneko, miya\}@ist.hokudai.ac.jp}

\maketitle

\begin{abstract}
Ensuring urban safety is an essential part of developing sustainable cities. An urban safety map can assist cities to prevent future crimes. However, mapping is costly in terms of both time and money due to the need for manual data collection. On the other hand, satellite imagery is becoming increasingly abundant and accessible with higher resolution. Given the outstanding success deep learning has achieved in the field of computer vision and pattern recognition over the past 5 years, in this paper we attempt to investigate the use of deep learning to predict crime rate directly from raw satellite imagery. To this end, we have trained a deep Convolutional Neural Network on satellite images obtained from over one million crime-incident reports collected by the Chicago Police Department. The best performing model predicted crime rate from raw satellite imagery with an accuracy of 79\%. To test their reusability, we used the learned Chicago models to predict for the cities of Denver and San Francisco city-scale maps indicating crime rate in three different levels. Compared to maps made from years' worth of data collected by the corresponding police departments, our maps have an accuracy of 72\% and 70\%, respectively.
\end{abstract}

\section{Introduction}
In a recent report \cite{UNODC2013}, the United Nations Office on Drugs and Crime (UNODC) has estimated that about a half million people worldwide are intentionally killed due to violence each year. With an average rate of 24 homicides per 100,000 inhabitants, countries in regions such as Latin America are particularly hit with endemic violence. Chronic violence has a significant cost (both direct and indirect) on the economy that in some countries reaches up to 10\% of GDP. This means that violence is not only a public health issue, but also a developmental one. Facing this grim situation, the world's nations have recently agreed to adopt the 2030 agenda of Sustainable Development which sets the goal of significantly reducing all forms of violence and related deaths everywhere by 2030 \cite{SDGs2015}.

An established practice for improving urban safety is to map crime incidents \cite{Chainey2013}. A map of previously committed crimes highlights where within the city criminal activities abound, and subsequently hints on where and what interventions are most needed. For example, increasing the frequency of police patrols around high-crime spots helps prevent future crimes.

Accurate mapping requires longitudinal data collection, which is highly costly in terms of both time and money. While rich countries are rich in data, most poor countries suffer from data poverty \cite{Leidig2016}. This is mainly due to the lack of means and/or the skills required for data collection. Therefore, an automatic approach to crime mapping is highly needed.

\begin{figure}[h]
\begin{tabular}{ccc}
 \includegraphics[scale=0.265]{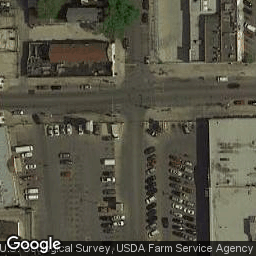} &
 \includegraphics[scale=0.265]{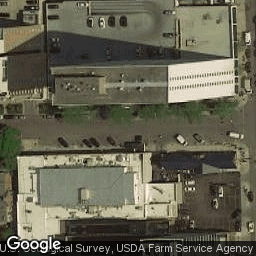} &
 \includegraphics[scale=0.265]{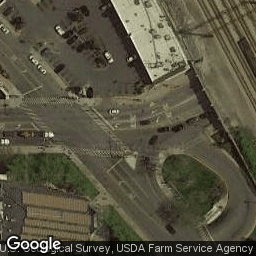} \\
 {(a)} &
 {(b)} &
 {(c)} \\
 \includegraphics[scale=0.265]{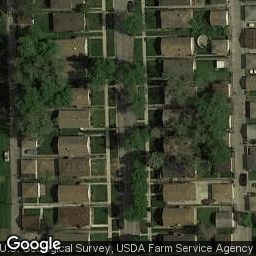} &
 \includegraphics[scale=0.265]{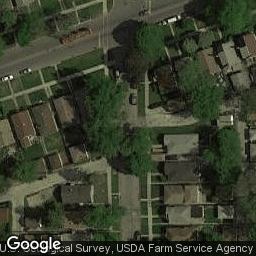} &
 \includegraphics[scale=0.265]{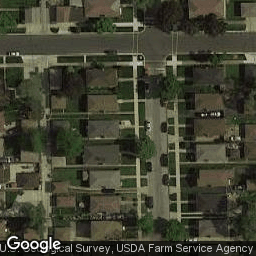} \\
 {(d)} &
 {(e)} &
 {(f)} \\ \\
\end{tabular}
\caption{Satellite images of six different locations in the city of Chicago. Between February 2012 and January 2016, there were over 100 crimes committed in each of the locations shown in the upper row. On the other hand and during the same period, there was only one crime committed in each of the locations of the bottom row. What is interesting is the high visual similarity among images of the same row. Notice how images of locations of similar crime rate have similar (1) setting (highway/parking lot vs. residential), (2) dominant color (gray vs. green), and (3) objects (road lines and vehicles vs. trees and rooftops). This example illustrates that visual features captured in satellite imagery can be used as a proxy indicator of crime rate. Therefore, based mainly on this observation we are motivated to investigate predicting crime rate directly from raw satellite imagery.}
\label{fig2}
\end{figure}

Recent progress in space and imaging technologies has made satellite imagery increasingly abundant and accessible with higher resolution \cite{Dash2016}. Satellite imagery has a bird's eye/aerial viewpoint which potentially makes it a rich medium of visual features relevant to different aspects of urban development. Given the outstanding success deep learning \cite{LeCun2015} has achieved in the field of computer vision and pattern recognition recently \cite{Donahue2014,Oquab2014,Razavian2014}, we are interested in investigating the use of deep learning to predict crime rate directly from raw satellite imagery.  

Satellite images pack a wealth of visual cues relevant to crime rate (See Figure \ref{fig2} for an illustrated example). However, satellite imagery is a highly \emph{unstructured} form of data. Moreover, given that learning robust computational models using current deep networks \cite{AlexNet2012} requires large numbers of training samples, it is difficult to extract useful insights on crime rate directly from raw satellite images using deep learning.

To overcome this problem we propose to leverage open government data which is both abundant and highly accurate. The idea is to mine massive numbers of crime-incident reports for high-quality training examples. These reports are collected by police departments over years and openly archived and updated online.

In this paper, we investigate the use of deep learning to accurately predict crime rate directly from raw satellite imagery. To this end, we have trained a deep Convolutional Neural Network (ConvNet) \cite{LeCun1989,LeCun1998} on satellite images obtained from over one million crime-incident reports collected over a period of 15 years by the Chicago Police Department. The best performing model predicted crime rate from raw satellite imagery with an accuracy of 79\%. To test their reusability, we used the learned Chicago models to predict for the cities of Denver and San Francisco city-scale maps indicating crime in three different levels. Compared to maps made from several years' worth of data collected by the corresponding police departments, maps predicted from raw satellite imagery have an accuracy of 72\% and 70\%, respectively.

To the best of our knowledge, this paper is the first to attempt predicting crime rate directly from satellite imagery. Contributions made in this paper are summarized bellow:
\begin{enumerate}
\item Presenting a proof-of-concept study on predicting crime rate from raw satellite imagery.
\item Proposing the idea of obtaining labeled satellite images from official crime-incident reports.
\item Predicting for the cities of Denver and San Francisco city-scale crime maps with an accuracy of 72\% and 70\%, respectively.
\end{enumerate}

The remainder of this paper is organized as follows. Previous works on machine learning-based urban safety mapping are briefly reviewed in section 2. Data used in this paper is described in section 3. Our approach of learning deep models for crime mapping is explained in section 4. Empirical validation using real data is presented in section~5. Finally, summary and conclusions are given in section~6.

\begin{table*}[htb]
\small
\centering
\begin{tabular}{>{\centering\arraybackslash} p{2cm}>{\centering\arraybackslash} p{2cm} >{\centering\arraybackslash} p{2cm} >{\centering\arraybackslash} p{2.5cm} >{\centering\arraybackslash} p{2.5cm} >{\centering\arraybackslash} p{2cm}} \noalign{\smallskip}\hline\noalign{\smallskip}
Report ID & Date  & Time  & Latitude & Longitude &  Category\\ \hline\noalign{\smallskip}
1  & 3/18/2016& 14:00&   41.xxxxxxx& -87.xxxxxxx& Arson\\
2  & 3/18/2016&  17:51&  41.xxxxxxx& -87.xxxxxxx& Homicide\\
3  & 7/06/2013&  15:00&  41.xxxxxxx& -87.xxxxxxx& Kidnapping\\
4  & 2/15/2016&  10:05&  41.xxxxxxx& -87.xxxxxxx& Robbery\\
5  & 2/04/2016&  21:50&  41.xxxxxxx& -87.xxxxxxx& Robbery\\
\noalign{\smallskip}\hline  \\
\end{tabular}
\caption{Examples of NIBRS-style crime-incident reports collected by the Chicago Police Department. Location information is anonymized for privacy concerns.}
\label{tab1}
\end{table*}

\section{Previous Works}
In this section, we briefly review previous works on urban safety (crime) mapping using machine learning, and compare them to ours.

To the best of our knowledge, the first major effort made at predicting city-scale urban safety maps is \cite{Naik2014}. Using an online crowdsourcing platform, a group of 7872 participants were shown random pairs of 4019 Google Street View images collected from the cities of New York, Boston, Salzburg, and Linz. For each pair, the participants were asked to choose the image they think looks safer. Then, individual images were assigned safety scores obtained from the accumulated preference vectors using the TrueSkill algorithm \cite{Herbrich2006}. Finally, each image was represented with a set of generic visual features collectively used to learn a model using Support Vector Regression \cite{Scholkopf2000}. The learned model was later used to generate city-scale safety maps for 27 US cities predicted directly from their Google Street View images. 

This study was recently extended in \cite{Dubey2016} to cover 29 more cities, using deep models learned from much larger pool of images annotated by over 81,000 participants.

Our work is similar to previous works in that both use visual information as a proxy indicator of urban safety. While~\cite{Dubey2016,Naik2014} use Google Street View Images, we use satellite imagery instead. 

On the other hand, the core difference between the two lies in the definition of urban safety. While in previous work, urban safety is subjectively judged by participants, we define urban safety based on the rate of crimes committed as reported by police departments.

Compared to ours, we believe that the mapping approach reported in \cite{Dubey2016,Naik2014} has the following limitations:
\begin{enumerate}
\item It is only viable in cities where services similar to Google Street View are available. Thus, it cannot be applied in most cities in developing countries. 
\item Building a robust model that can predict urban safety from natural images requires crowdsourcing the votes of tens of thousands of online participants, a process that is both time consuming and labor intensive.
\end{enumerate}

We are aware of other works, such as \cite{Kianmehr2008}, which mainly focus on the prediction of crime-prone areas (crime hotspots) rather than the prediction of crime at a city scale. Therefore, and given the above, we believe that our work is the first to attempt using machine learning to predict city-scale crime maps directly from satellite imagery.

\section{Data}
In this section, we describe the data we used to train, verify, and test our deep models. 

All models in this study were trained, verified and tested on satellite images mined from official crime-incident reports collected by police departments in the United States. These reports are released as \emph{open data} which is defined as data that can be freely used, re-used and redistributed by anyone - subject only, at most, to the requirement to attribute and sharealike \cite{OD2009}. 

Used reports follow the National Incident Based Reporting System (NIBRS) \cite{Maxfield1999} which describes individual crime incidents using several attributes, such as time, date, (approximate) geographic location, and category of crime. See Table~\ref{tab1} for examples.

We used crime reports collected in three US cities: Chicago, Denver, and San Francisco. Collected data include both violent and nonviolent crimes, such as bribery and fraud. However since we are interested in predicting violent crimes, we discarded all non-violent entries. The used data can be summarized as follows:
\begin{itemize}
\item 1,028,885 crime reports collected by the Chicago Police Department over the period between September 2001 and August 2016 \cite{CPD2016}. 
\item 198,506 crime reports collected by the Denver city police department over the period between July 2014 and July 2016 \cite{DPD2016}. 
\item 652,807 crime reports collected by the San Francisco police department over the period between March 2003 and September 2016 \cite{SFPD2016}.
\end{itemize}

Given that police reports were collected manually by skilled enumerators, we assume that models learned from this data are highly accurate. Using the geographic location information, we represented individual reports with satellite images. We used these images to train, verify, and test our computational models.

\section{Deep Models for Crime Mapping}
At first glance, the ability to map crime might imply the ability to predict the occurrence of individual criminal incidents. However, since crimes are committed as a result of a complex combination of social, economical, etc. factors, the task of predicting individual crime incidents is extremely challenging, and beyond the interest of this paper. Instead, we are interested in generating city-scale maps~\cite{Chainey2002} that indicate crime rate in three different levels: low, neutral, and high. Assuming that satellite imagery is rich in visual cues that can be effectively used as a proxy indicator of crime rate, we aim to predict our maps directly from raw satellite images.

In the following, we describe our approach. First, we explain how we mine official police reports for satellite images labeled with crime rate. Then, we describe how we use the obtained data to train deep models able to predict crime rate directly from satellite imagery.

\subsection{From Police Reports to Satellite Images}
Learning a robust deep model requires collecting a set of accurately labeled training images, of a scale large enough to train a deep network. In the same spirit of \cite{Najjar2017}, to this end we propose to mine open geotagged police reports. This procedure is explained in the three steps below:

\subsubsection{Location information discretization}
First, we divided the satellite map of the city of Chicago using a square grid into equal regions ($r$) of 900 square meters each. Then, given their location information (latitude longitude pairs), we assigned the 1,028,885 crimes collected by the Chicago police department to their corresponding regions accordingly. Finally, we assigned each region a safety score~($S_{r}$) given as the number of  crimes committed within its boundaries, such that:
\begin{equation}
S_{r} = \sum_{i=1}^{n}{a_{i,r}},
\end{equation}
\noindent where $a_{i, r}$ is the $i$-th crime committed within region $r$, and $n$ is the total number of crimes.

\subsubsection{Binning}
Since the goal is to generate maps that indicate crime rate in three levels, training examples need to be categorized into three safety levels (low, neutral, and high). To this end, we clustered the obtained safety scores by frequency around three bins using the k-means algorithm \cite{Macqueen1967}, such that:
\begin{equation}
\arg \min_{T} \sum_{i=1}^{k} \sum_{x \in T_{i}}{\|x-\mu_{i}\|^{2}},
\end{equation}

\noindent where $\mu_{i}$ is the mean of the points in $T_{i}$, $k = 3$ is the number of bins, and $x$ is the frequency of individual scores. We have tried other clustering algorithms, such as Jenks natural breaks optimization \cite{Jenks1967} and Gaussian Mixture Models (GMM). However, we found that k-means delivers the best results.

\subsubsection{Resampling}
Since the obtained labeled set of regions is highly imbalanced. Therefore and in order to avoid learning a biased model, we resampled our data via downsampling majority classes so that the three classes are balanced out.

Finally, we represented each square region with a satellite image centered around the GPS coordinates of its center.

\subsection{Crime Rate Prediction using ConvNets}
We begin by introducing ConvNets, then we explain how our models are learned.

\subsubsection{Convolutional Neural Networks}
Designed to process data in multiple array format, a Convolutional Neural Network (ConvNet) is a deep feedforward neural network that automatically learns from training data hierarchical representations capturing patterns and statistics at multiple levels of abstraction. A typical ConvNet consists of a set of convolutional layers followed by a set fully-connected layers ordered such that the output of one layer is the input to the next. As an input a ConvNet takes a multi-dimensional array (e.g., RGB image), and produces a class/label prediction as an output. A ConvNet is trained end-to-end in a supervised fashion using Stochastic Gradient Descent (SGD) and back-propagation. 

For more information on ConvNets and how they work, the reader is referred to \cite{LeCun2015}. 

\subsubsection{Model Learning}
We adopted transfer learning in training our models. In transfer learning, a pre-learned knowledge is transferred from a source to a target problem. In our case, source and target problems are generic object/scene recognition, and crime rate prediction, respectively. The transferred knowledge is a set of low- and mid-level visual features such as edges, corners and textures. This method of training is commonly referred to as finetuning in the deep learning community, and it has been proven highly successful in enhancing learning when training examples are scarce \cite{Branson2014,Karayev2013}.
 
To finetune a pre-trained model, we first replaced the source classification layer with a three-class output layer~(representing the three safety labels). Weights of the new layer are randomly initialized, and finally the entire network is trained jointly using small learning rates.

\begin{table*}[htb]
\small
\centering
\begin{tabular}{>{\centering\arraybackslash} p{4cm}>{\centering\arraybackslash} p{2cm} >{\centering\arraybackslash} p{2cm} >{\centering\arraybackslash} p{2cm} >{\centering\arraybackslash} p{2cm}} \noalign{\smallskip}\hline\noalign{\smallskip}
                    & x17   & x18  & x19  & x20 \\\hline\noalign{\smallskip}
ImageNet            & 0.763 & 0.727 & 0.702 & 0.643 \\
Places205           & \textbf{0.795} & 0.748 & 0.728 & 0.638 \\
ImageNet + Places205& 0.782 & 0.733  & 0.725 & 0.673 \\
\noalign{\smallskip}\hline  \\
\end{tabular}
\caption{Average prediction accuracy obtained using different models pre-trained on three different large-scale datasets and finetuned on satellite images captured at four different zoom levels. The best performing model is the one pre-trained on Places205 and finetuned on satellite images captured at zoom level 17. All results are cross-validated on three random data splits.}
\label{tab2}
\end{table*}

\section{Experiments}
In this section, we empirically validate our deep models by evaluating the accuracy of crime maps predicted for two cities: (1) Denver, and (2) San Francisco using models learned from data collected in Chicago.

\subsection{Datasets}
In our experiments, we used three different satellite imagery datasets collected and labeled as detailed in the previous section:

\emph{Chicago}: 12,000 satellite images obtained from official crime reports collected by the Chicago Police Department.

\emph{Denver}: 25,169 satellite images obtained from official crime reports collected by the Denver city Police Department.

\emph{San Francisco}: 19,897 satellite images obtained from official crime reports collected by the San Francisco Police Department.

\subsection{Implementation Details}

We implemented our experiments as described below: 

\emph{Image data}: we used Google Static Maps API \cite{Maps2016}, to crawl all satellite imagery. Spatial resolution is set to 256$\times$256 pixels each, spanning four different zoom levels (17 through 20).

\emph{Deep architecture}: we adopted AlexNet architecture as described in \cite{AlexNet2012}. We are aware of the other successful deep architectures, such as \cite{Simonyan2014,GoogLeNet2015}. However, we settled on AlexNet since it is both simple and considered a landmark classifier.

\emph{Training}: we initialized all models from generic large-scale image datasets. Three datasets were considered: (1)~ImageNet \cite{ImageNet2009}, (2) Places205 \cite{Zhou2014}, and (3) both ImageNet and Places205 combined. We performed training using Caffe framework \cite{Caffe2014} run on a single Nvidia GeForce TITAN X GPU.

\emph{Evaluation}: to evaluate the learned models, we used the average prediction accuracy (cross-validated on three random 5\%/95\% data splits) as an evaluation metric. Results were obtained after 60,000 training iterations.

\subsection{Predicting Crime Rate in Chicago}
The purpose of this experiment is twofold: (1) to investigate whether or not visual features captured in satellite imagery can be effectively used as a proxy indicator of crime rate. And (2) to evaluate the performance of state-of-the-art ConvNets in learning deep models able to predict crime rate from raw satellite images.

The result of finetuning on \emph{Chicago} dataset is shown in Table \ref{tab2}. This table shows average prediction accuracy of 12 models obtained considering three pre-training datasets and four finetuning scenarios. 

Ranging between 63.8\% and 79.5\%, the best performing model is the one obtained through finetuning a pre-trained model on Places205 dataset using satellite images captured at zoom level 17. 

From Table \ref{tab2}, we make the following observations:
\begin{enumerate}
\item For all zoom levels (except zoom level 20), models pre-trained on Places205 perform the best, followed by models pre-trained on both Places205 and ImageNet, and finally models pre-trained on ImageNet. This is expected since satellite images have bird's eye/aerial viewpoint which makes them closer in nature to scene images of Places 205 rather than the object-centric images of ImageNet.
\item For all pre-training scenarios, models finetuned using satellite images captured at zoom level 17 perform the best. On the other hand, models finetuned on zoom level 20 images perform the worst.
\end{enumerate}

Results obtained in this experiment confirm our assumption that visual features captured in satellite imagery can be efficiently used as a proxy indicator of crime rate. Moreover, state-of-the-art ConvNets are able to learn robust models that can predict crime rate from raw satellite images.

\subsection{City-Scale Crime Mapping}
The purpose of this experiment is to empirically evaluate the reusability of the learned deep models. To this end, we applied Chicago models to generate city-scale crime maps predicted from raw satellite imagery for two US cities, namely Denver and San Francisco.

\begin{figure*}[htb]
\centering
\begin{tabular}{cccc}
 \includegraphics[scale=0.17]{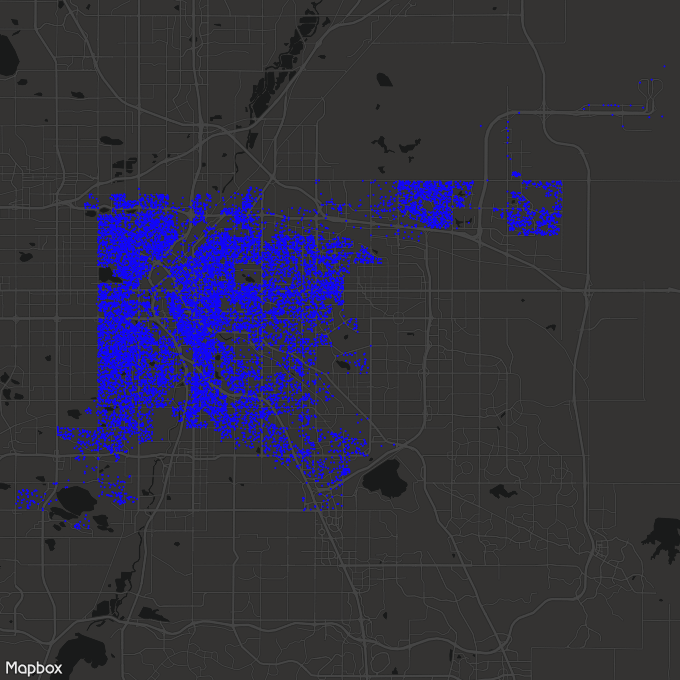} &
 \includegraphics[scale=0.17]{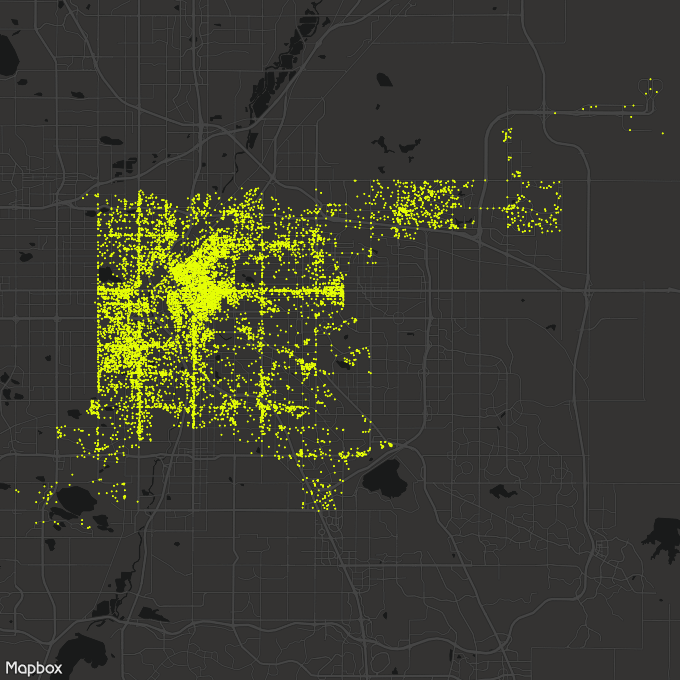} &
 \includegraphics[scale=0.17]{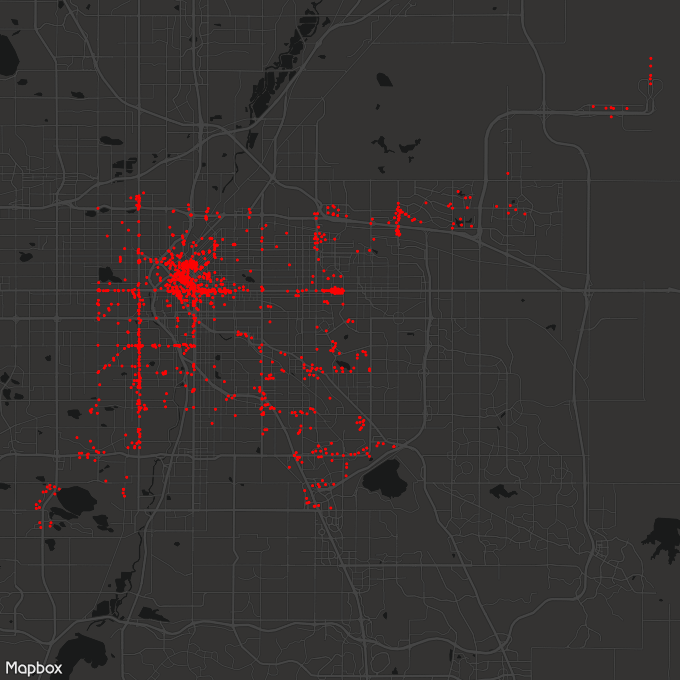} &
 \includegraphics[scale=0.17]{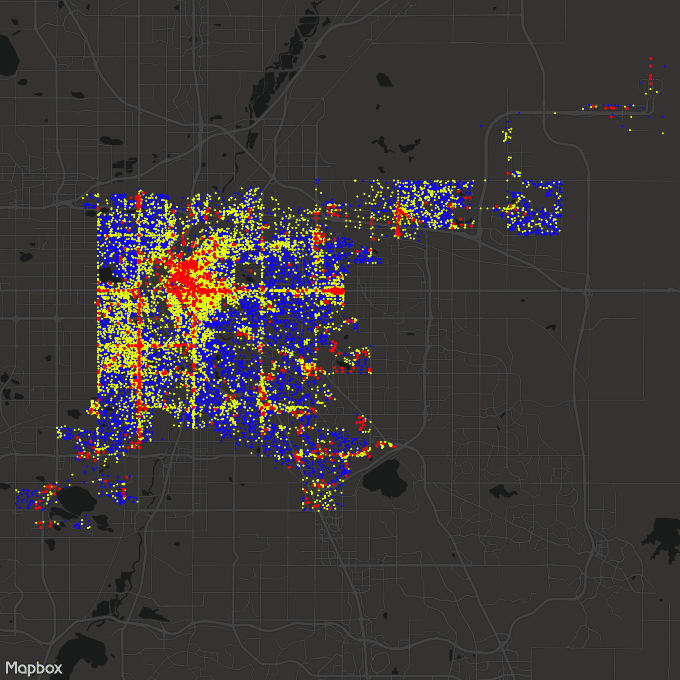} \\
 {(a)} &
 {(b)} &
 {(c)} &
 {(d)} \\
  \includegraphics[scale=0.17]{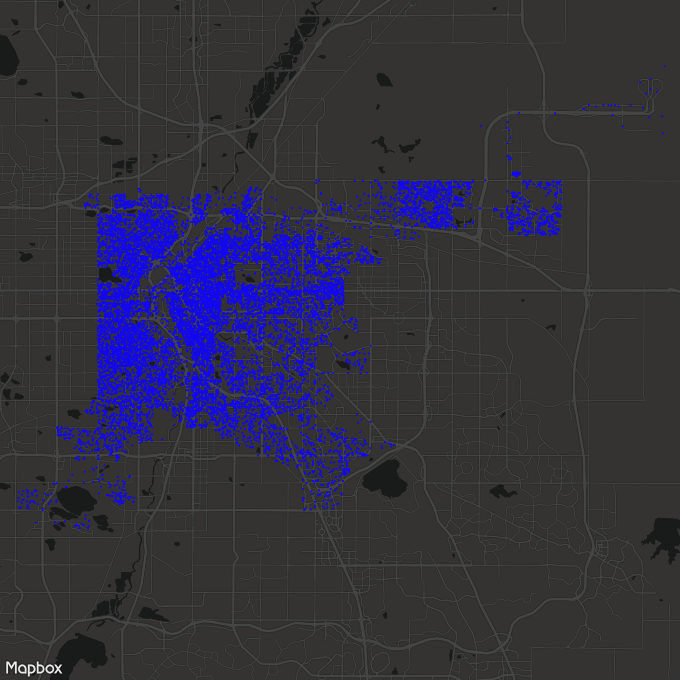} &
 \includegraphics[scale=0.17]{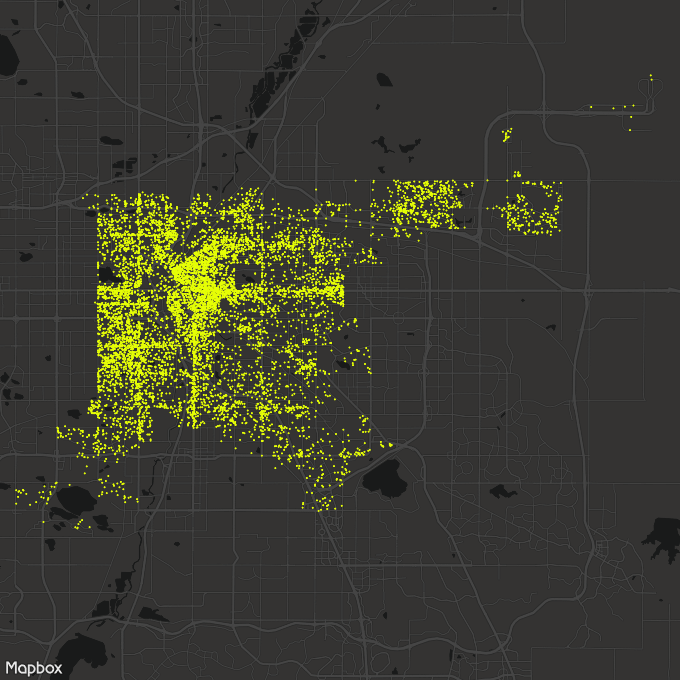} &
 \includegraphics[scale=0.17]{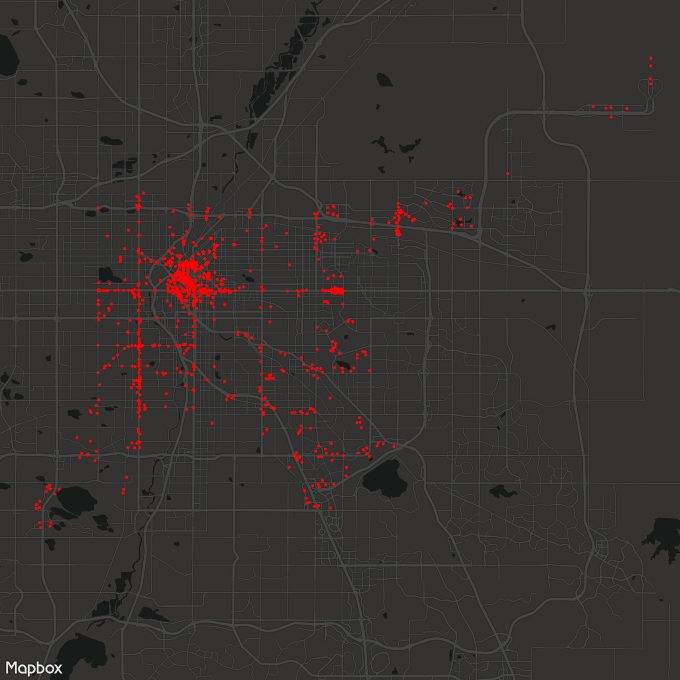} &
 \includegraphics[scale=0.17]{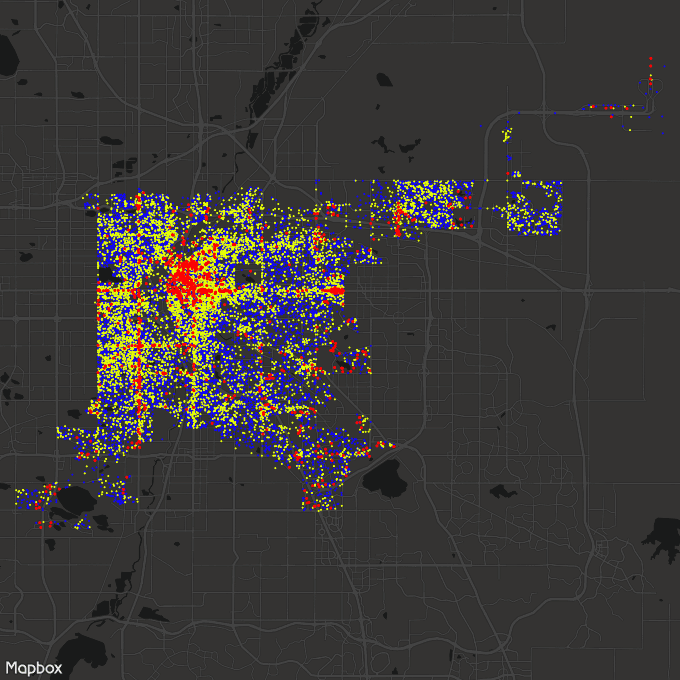} \\
 {(d)} &
 {(e)} &
 {(f)} &
 {(g) 72.7\%} \\ \\ \\
 \includegraphics[scale=0.25]{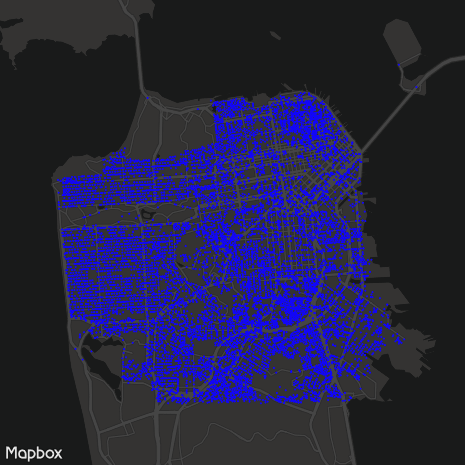} &
 \includegraphics[scale=0.25]{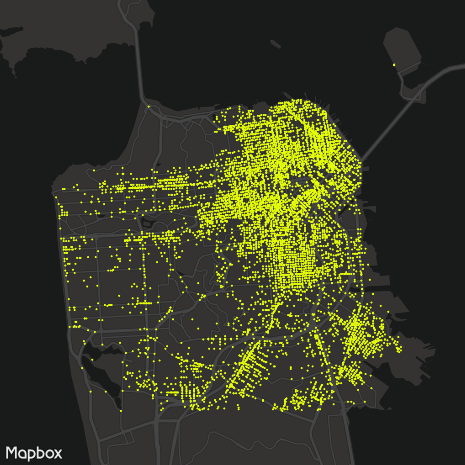} &
 \includegraphics[scale=0.25]{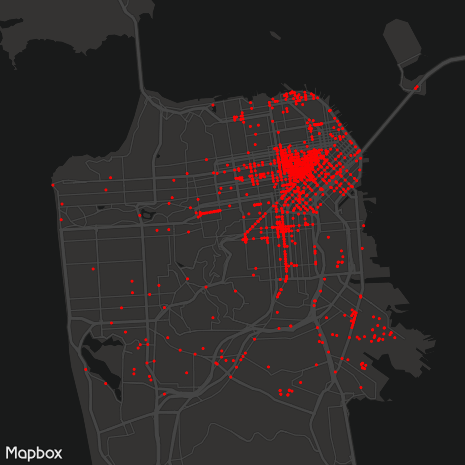} &
 \includegraphics[scale=0.25]{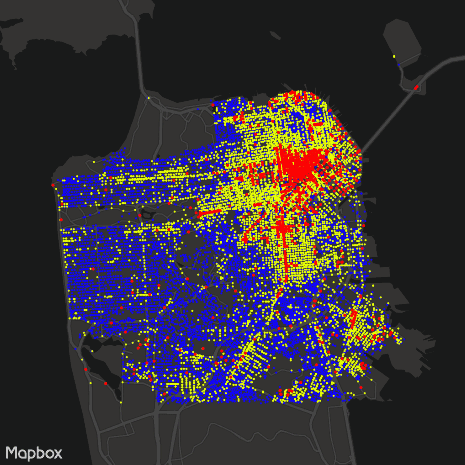} \\
 {(h)} &
 {(i)} &
 {(j)} &
 {(k)} \\ 
 \includegraphics[scale=0.25]{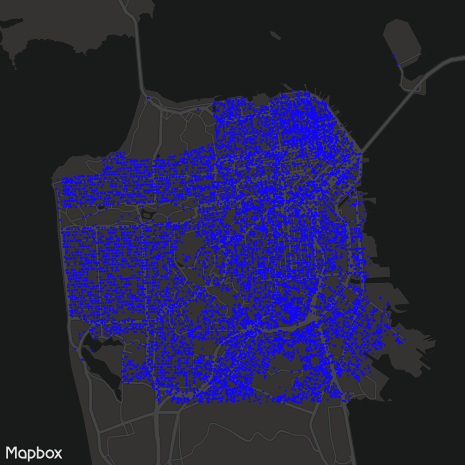} &
 \includegraphics[scale=0.25]{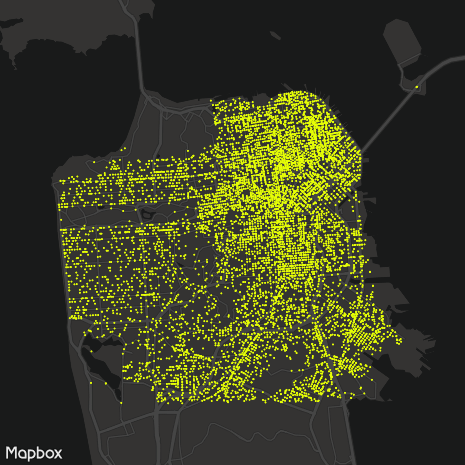} &
 \includegraphics[scale=0.25]{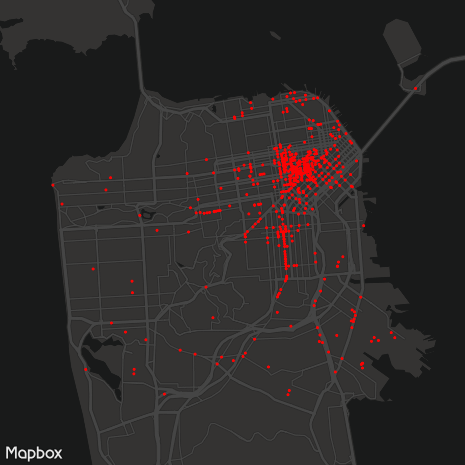} &
 \includegraphics[scale=0.25]{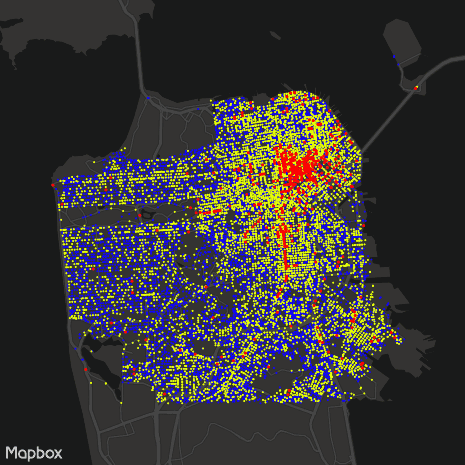} \\
 {(l)} &
 {(m)} &
 {(n)} &
 {(o) 70.8\%} \\ \\ \\
\end{tabular}
\caption{City-scale crime maps of the cities of Denver (a-g) and San Francisco (h-o). For each city, the upper row is a map made from official data. While, the bottom row is a map completely predicted from raw satellite imagery. First three columns (left to right) represent the three crime rate labels (low: blue, neutral: yellow, and high: red) mapped individually. The fourth column represents all labels mapped together. The predicted maps have an accuracy of 72.7\% and 70.8\%, respectively. Best viewed in digital format.}
\label{fig3}
\end{figure*}

We used the best performing Chicago models to predict labels of \emph{Denver} dataset images. Figure~\ref{fig3}(a-g) shows a city-scale crime map for the city of Denver. The upper row is a map made from official crime reports collected by the Denver police department over the period between July 2014 and July 2016. The bottom row shows a map predicted completely from raw satellite images. Compared to the official map (upper row), the predicted map (bottom row) has an accuracy of 72.7\%. 

We also predicted safety labels of \emph{San Francisco} dataset images. Figure \ref{fig3}(h-o) shows a city-scale crime map for the city of San Francisco. The upper row is a map made from official crime reports collected by the San Francisco police department over the period between March 2003 and September 2016. The bottom row shows a map predicted completely from raw satellite images. Compared to the official map (upper row), the predicted map (bottom row) has an accuracy of 70.8\%.

For both maps, the first three columns (left to right) illustrate the three crime rate labels (low: blue, neutral: yellow, and high: red) mapped individually. The fourth column illustrates the three labels mapped together.

Since Chicago is quiet different from both Denver and San Francisco in terms of population, area, and crime rate, demonstrating that a model learned from data collected in Chicago can effectively (to a certain degree) predict crime rate in both Denver and San Francisco proves that our learned models are practically reusable. Moreover, in order to quantify the accuracy of the predicted maps, we had to choose cities that have their official crime data publicly accessible so that we can compare our results to a ground truth. On the basis of these criteria we have decided to map the cities of Denver and San Francisco in this experiment.

Results obtained in this experiment confirm that deep models learned from crime data collected in one city can be reused in different cities.

\section{Summary \& Conclusions}
In this paper, we have attempted to investigate the use of deep learning to predict city-scale crime maps directly from satellite imagery. Our approach to mapping uses Convolutional Neural Networks (ConvNets) and leverages open governmental data to learn robust deep models able to predict crime rate from raw satellite imagery.

We have trained deep models on satellite images obtained from over one million crime reports collected over a period of 15 years by the Chicago Police Department. The best model predicted crime rate from raw satellite imagery with an accuracy of 79\%. We also used the learned models to predict for both cities of Denver and San Francisco city-scale maps indicating crime rate in three levels. Compared to maps made from years' worth of official data, our maps have an accuracy of 72\% and 70\%, respectively. 

The obtained results confirm that: (1) visual features contained in satellite imagery can be effectively used as a proxy indicator of crime rate. (2) State-of-the-art ConvNets are able to learn robust models for crime rate prediction from satellite imagery. (3) Learned deep models can be reused across different cities.

Although providing a proof-of-concept study on crime mapping where proper data collection is not affordable~(e.g., poor countries), our work suffers from several limitations. First, our models do not take crime category into consideration. We have used crime incident count only as image labels. We believe that training models on more elaborate data will result in more insightful maps. Second, our models predict crime rate without taking time into consideration. In other words, our maps do not differentiate between day and night or summer and winter. Third, although we proved our approach effective (to a certain degree) in predicting crime rate in Denver and San Francisco using models trained on data collected in Chicago, we have not considered a more extreme case in which both cities are located in two different continents (e.g., Chicago and Nairobi) where architecture, city planning, level of development, etc. differ extremely. These limitations among others are to be addressed in future work.

{\small
\bibliographystyle{ieee}
\bibliography{mybib}
}

\end{document}